\documentclass[prd,aps,reprint,onecolumn,superscriptaddress,tightenlines,nofootinbib,eqsecnum,preprintnumbers,longbibliography,12pt]{revtex4-2}
\pdfoutput=1
\usepackage[T1]{fontenc} 
\usepackage[dvipsnames]{xcolor}
\usepackage{bbm}
\usepackage{amsmath,amssymb}
\usepackage{physics}
\usepackage{dsfont}
\usepackage{graphicx}
\usepackage{hyperref}
\usepackage{slashed}
\usepackage{comment}
\usepackage{subcaption} 
\usepackage{adjustbox}  
\usepackage{soul}
\DeclareUnicodeCharacter{2212}{-}
\newcommand{\be}{\begin{equation}}
\newcommand{\ee}{\end{equation}}
\newcommand{\bea}{\begin{eqnarray}}
\newcommand{\eea}{\end{eqnarray}}
\newcommand{\nn}{\nonumber}

\begin{document}


\begin{flushright}
 CP3-22-40 
\end{flushright}

\title{P-wave Sommerfeld enhancement near threshold: a simplified approach
}

\author{Carlos H. de Lima}
\email{CarlosHenriquedeLima@cmail.carleton.ca}
\affiliation{Ottawa-Carleton Institute for Physics, Carleton University, \\ Ottawa, ON K1S 5B6, Canada}
\author{Alberto Tonero}
\email{alberto.tonero@gmail.com}
\affiliation{Ottawa-Carleton Institute for Physics, Carleton University, \\ Ottawa, ON K1S 5B6, Canada}
\author{Andres Vasquez}
\email{andres.vasquez@uclouvain.be}
\affiliation{Centre for Cosmology,Particle Physics and Phenomenology (CP3),\\ Universit\'e Catholique de Louvain, B-1348 Louvain-la-Neuve, Belgium}
\affiliation{ICTP South American Institute for Fundamental Research \& Instituto de F\'isica Te\'orica \\ Universidade Estadual Paulista \\ Rua Dr. Bento T. Ferraz 271 - 01140-070 S\~ao Paulo, SP, Brazil S\~ao Paulo, Brazil}
\author{Rogerio Rosenfeld}
\email{rogerio.rosenfeld@unesp.br}
\affiliation{ICTP South American Institute for Fundamental Research \& Instituto de F\'isica Te\'orica \\ Universidade Estadual Paulista \\ Rua Dr. Bento T. Ferraz 271 - 01140-070 S\~ao Paulo, SP, Brazil S\~ao Paulo, Brazil}

\date{\today}
\begin{abstract}
The calculation of P-wave Sommerfeld enhancement in processes with unstable particles in the final state is known to be divergent. In a complete description, where resonant (on-shell unstable particles) 
and non-resonant contributions are included, it has been shown that results are finite.  For most beyond the Standard Model applications, these complete calculations are not readily available.
In this work, we are interested in the near-threshold region and we consider only the resonant contribution.  In this case, we provide a simplified prescription to compute the P-wave Sommerfeld enhancement in the narrow-width approximation of the unstable particle that directly eliminates divergences. We show that we can define a finite resonant contribution without the inclusion of the non-resonant processes in a way similar to the usual S-wave Sommerfeld enhancement.

\end{abstract}

\maketitle

\section{Introduction and motivation}
\label{sec:intro}

The Sommerfeld enhancement \cite{SOMM} in scattering amplitudes can be understood as the result of multiple exchanges of light mediators among either the initial or the final state particles in the process. This is a non-relativistic quantum mechanical effect that can be expressed as a deformation of the initial or final state wavefunctions induced by the interaction of the external states with the light mediators. This modification can be obtained by solving the Schr\"odinger equation with a potential term that results from the new interaction. Typically, these effects lead to significant enhancements in the cross-sections near the threshold due to non-perturbative effects. These effects can overcome phase space suppression, resulting in a finite cross-section even at the threshold. 

The first computation of the Sommerfeld enhancement due to the Coulomb interaction between final-state leptons was performed in \cite{Sakharov:1948plh}. Examples of Sommerfeld enhancement computations in S-wave dominated processes involving stable particles can be found in studies related to dark matter annihilation~\cite{Hisano:2003ec,Hisano:2004ds,Arkani_Hamed_2009,Lattanzi:2008qa}. General results for arbitrary partial waves have been obtained in~\cite{Iengo:2009ni, Cassel:2009wt}. Examples of Sommerfeld enhancement computations in S-wave dominated processes involving unstable particles can be found in studies related to top quark pair production~\cite{topFadin,Fadin:1990wx,Peskin,toppair,Jezabek:1998pj}, $W^+ W^-$ production~\cite{WFadin} and Higgs pair production~\cite{Grinstein:2007iv,Oliveira:2010uv}. Sommerfeld enhancement studies for processes that have significant P-wave contributions and involve unstable particles (like stop quarks) in the final states, have been performed in~\cite{stopFadin,Fabiano:2001cw,pwaveTH,KIM2019469}.

The computation of the Sommerfeld enhancement for unstable particles in the final state uses Green's functions and the optical theorem. On one hand, the presence of a finite width in the propagator regulates an infrared divergence by damping the wave functions at large distances. On the other hand, it also introduces ultraviolet divergences in the computation of the Sommerfeld enhancement for P-wave dominated processes \cite{stopFadin,Penin:1998mx,pwaveTH}. The reason is that, in this case, the enhancement is obtained from the Laplacian of the Green's function, which is divergent at the origin (short distance). This divergence is real when dealing with stable particles or considering S-wave dominated processes with unstable particles, and hence it does not contribute to the cross-section. When considering  P-wave dominated processes with unstable particles, the width introduces a divergence in the imaginary part of the amplitude that contributes to the cross-section through the optical theorem. 

 These divergences are usually resolved by considering the physical process in the complete theory which includes the non-resonant contributions\cite{Beneke:2003xh,Beneke:2004km,Jantzen:2013gpa}. Using this approach, if one is interested in calculating a P-wave process with unstable particles, one needs to compute the all-orders effect from the Sommerfeld enhancement and also the leading perturbative process at a higher loop. The calculation of the perturbative higher loop is not readily available for most processes. Furthermore, it accounts for a small correction of the finite resonant piece above threshold\footnote{For instance, it accounts for {\cal O}(5\%) in the $e^{+}e^{-}\rightarrow t \bar{t}$ process at NLO \cite{Jantzen:2013gpa}.}. 
In this work, we develop a method to obtain the finite resonant corrections without having to compute the higher-loop non-resonant process. This is useful for beyond the standard model processes to obtain an estimation of P-wave processes with unstable particles which are enhanced by the exchange of a light mediator. In this case, the computation of the leading resonant contribution is sufficient to get a good estimate of the cross-section~\cite{Jantzen:2013gpa} including the width effects.
Focusing only on the threshold region,  we show that the resonant contribution of a Sommerfeld enhancement can be computed  in the non-relativistic quantum field theory with unstable particles assuming a narrow width approximation~\footnote{
Unstable particles have to be treated with care when considered as virtual particles in the optical theorem. The validity of the optical theorem for unstable particles is discussed in Appendix \ref{ap:OPT}.}.

The remainder of this paper is organized as follows. In section~\ref{sec:sommerfeld}, we briefly review the Sommerfeld enhancement effect in a 2-to-2 process due to the exchange of a light mediator in the final state. We consider the annihilation of a fermion-antifermion into two scalars as a concrete example. In section~\ref{sec:Scalc} we introduce the general formalism for the calculation of the enhancement in the presence of an unstable particle in the final state using the optical theorem. In section~\ref{sec:warm} we present the computation of the enhancement for a specific S-wave dominated process and we propose a procedure to deal with the UV divergences appearing in the calculation. 

In section~\ref{sec:pwave} we present the computation of the enhancement for a specific P-wave dominated process. In this case, the divergences are imaginary and by applying our method we show how the renormalization procedure provides a finite unambiguous result. We discuss some of the differences between using our renormalization procedure and a finite cutoff in section~\ref{sec:discuss}. We conclude in section~\ref{sec:CONC}. In appendix \ref{ap:nonrel} we derive the non-relativistic approximation of the recursion relations used in the main text and in Appendix \ref{ap:OPT}  we discuss the validity of the optical theorem for unstable particles.

\section{Brief Review of Sommerfeld enhancement } \label{sec:sommerfeld}
The Sommerfeld enhancement for a 2-to-2 scattering process is due to the exchange of a light force carrier between the initial or final state particles and can be computed using two distinct methods. The first method, which is the way the effect was discovered, uses the computation of the wave function of the produced final state  \cite{Lattanzi:2008qa, DMSOMM, Iengo:2009ni, Cassel:2009wt}. The calculation of this wave function at the origin includes the non-perturbative information of the interaction potential and gives the enhancement factor of the scattering process. This approach is straightforward and uses basic quantum mechanics. The second method, instead, takes advantage of the optical theorem to relate the cross-section with the imaginary part of an amplitude where initial and final states are the same \cite{Peskin, toppair, topFadin}. In this work, we adopt this second approach to use the full power of quantum field theory and tackle the divergences that arise in the computations.

\begin{figure}[h!]
 \resizebox{0.8\linewidth}{!}{\includegraphics{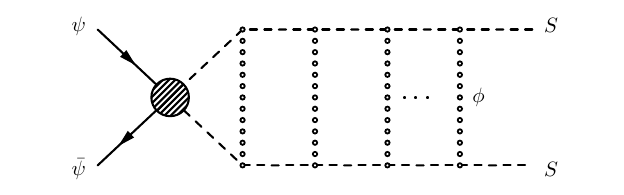}}
  \caption{\label{fig:Sommer_sketch} Sommerfeld enhancement in the process $\psi\bar{\psi} \to SS$ due to the mediation of the scalar $\phi$.}
\end{figure}

To demonstrate our methodology, in this paper we consider the annihilation of two fermionic particles $\psi$ into two massive scalars $S$ with mass $m_S$ and width $\Gamma_S$~\footnote{In principle, the field $S$ can be a complex scalar but for simplicity, we take it to be real.}. We will further assume the existence of a light real scalar field $\phi$ which interacts exclusively with $S$ in such a way that the cross-section of the process $\psi  \bar{\psi} \to S S$ can be enhanced by the multiple exchanges of $\phi$-particles between the two scalar particles in the final state, as shown in Fig. \ref{fig:Sommer_sketch}. In the non-relativistic limit, the enhancement is dominated by the ladder diagrams which are related to a threshold singularity. Each $\phi$ exchange generates an enhancement factor of ($\alpha_{\phi}/v$). At this point, we need to systematically resum the ladder diagrams or use the wave function under the Yukawa potential in order to get the precise
annihilation cross-section.  Let us define the enhancement of the cross-section in this process as:

\begin{align}
 \sigma (\psi  \bar{\psi} \to S S)={\cal S}(E)\,\sigma_0 (\psi  \bar{\psi} \to S S)    \, ,
\end{align}
where $\sigma_0 (\psi  \bar{\psi} \to S S)$ is the leading order cross-section and ${\cal S}(E)$ is the energy-dependent Sommerfeld enhancement factor. The computation of the enhancement factor ${\cal S}(E)$ depends on two important elements of the theory. The first one is the form of the $\phi S S$ interaction vertex. It can be generic, but in this work, we assume a simple form that occurs in a plethora of different models, namely:
\begin{align}
    \mathcal{L}_{\rm int}  = \frac{\kappa}{2} \phi S^{2}  \, ,
\end{align}
where $\kappa$ is a dimensionful coupling constant. The second element is the leading order amplitude ${\cal M}_0(\psi  \bar{\psi} \to S S)$ such that:
\begin{align}
\sigma_0 (\psi  \bar{\psi} \to S S) \sim |{\cal M}_0(\psi  \bar{\psi} \to S S)|^2.
\end{align}
The most general leading-order (off-shell) amplitude for the process $\psi(q_1)  \bar{\psi}(q_2) \to S(p_1) S(p_2)$ can be written as:
\begin{align} \label{eq:ampbase}
   {\cal M}_0(\psi  \bar{\psi} \to S S) = \bar v (q_2) \Gamma_I u(q_1) F_0^I(q,p) \, , 
\end{align}
where $q=q_1+q_2=p_1+p_2=\sqrt{s}$ is the center of mass energy and $\Gamma_I=1,\gamma_\mu,\gamma_5,\gamma_\mu\gamma_5,\sigma_{\mu\nu}$ is an element of the Clifford space basis. By defining $p_1=q/2+p$, $p_2=q/2-p$ and using momentum conservation, we have that the leading order form factor  $F_0^I$ is a function of three independent momenta: $q_1$, $q$ and $p$. For simplicity, we just write $F_0^I(q,p)$.
\begin{figure}
\centering
\begin{minipage}[t]{0.32\textwidth}
    \begin{subfigure}{\textwidth}
        \centering
        \scalebox{0.8}{\includegraphics{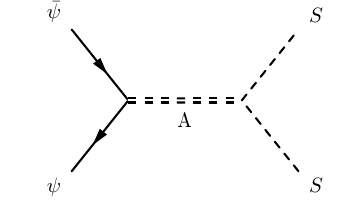}}
        \caption{}
        \label{fig:case1}
    \end{subfigure}
\end{minipage}
\begin{minipage}[t]{0.32\textwidth}
    \begin{subfigure}{\textwidth}
        \centering
        \scalebox{0.8}{\includegraphics{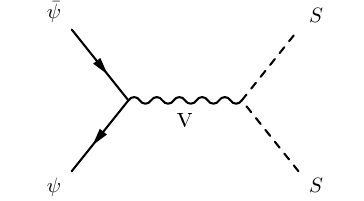}}
        \caption{}
        \label{fig:case2}
    \end{subfigure}
\end{minipage}
\begin{minipage}[t]{0.32\textwidth}
    \begin{subfigure}{\textwidth}
        \centering
        \scalebox{0.8}{\includegraphics{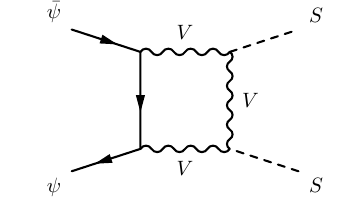}}
        \caption{}
        \label{fig:case3}
    \end{subfigure}
\end{minipage}
\caption{Examples of processes that are dominated by a single partial-wave: (a) S-wave dominated, (b) P-wave dominated at tree-level and (c) P-wave dominated at 1-loop.}
\label{fig:cases}
\end{figure}

At this point, we can expand the amplitude in partial waves as follows
\be 
{\cal M}_0 = 16\pi \sum_{l=0}^\infty (2l+1) P_l(\cos\theta){\cal M}_0^l(s,t),
\ee
where $s,t,u$ are the usual Mandelstam variables, $P_l(x)$ is the Legendre polynomial of order $l$ and ${\cal M}_0^l$ is the $l$-th partial wave amplitude. Even though this expansion is not a necessary step for performing the Sommerfeld enhancement computation, it represents a very useful tool that simplifies the study of the enhancement in cases where only one partial wave dominates. 

For instance, S-wave $(l=0)$ amplitudes can arise in renormalizable models where the $\psi \bar{\psi} \to S S$ process is mediated by the exchange of a scalar particle $A$ with mass $m_A$ in the s-channel (see Fig. \ref{fig:case1}) and they can be written as: 
\begin{align} \label{eq:ampS}
{\cal M}^{\rm S-wave}_0(\psi  \bar{\psi} \to S S) \propto \frac{y_{\psi A}\kappa_{SA}}{q^2 - m_{A}^{2}}\bar v (q_2) u(q_1)   \, ,
\end{align}
where $y_{\psi A}$ is the $\bar \psi \psi A$ Yukawa coupling and $\kappa_{SA}$ is the trilinear $ASS$ scalar coupling.
One concrete occurrence of such amplitude is the tree-level Higgs pair production at muon colliders, which is dominated by the s-channel exchange of the Higgs itself and $y_{\psi A}$ is the lepton Yukawa coupling and $\kappa_{SA}$ is the triple Higgs coupling $\lambda_h$.

An example of P-wave $(l=1)$ amplitude can arise in renormalizable models where we have a vector field $V_\mu$ with mass $m_V$ which couples to $\psi$ and $S$~\footnote{Notice that in this case $S$ is charged and the final state will be $S\bar S$.}. In this case, the $\psi \bar{\psi} \to S  S$ process is mediated by the exchange of $V_\mu$ in the s-channel (see Fig. \ref{fig:case2}) and the amplitude can be written as:
\begin{align}\label{eq:ampP}
{\cal M}^{\rm P-wave}_0(\psi  \bar{\psi} \to S  S) \propto \frac{g_{\psi }g_{S}}{q^2 - m_{V}^{2}}\bar v (q_2)\slashed{p}\, u(q_1)    \, ,
\end{align}
where $g_\psi$ and $g_S$ are, respectively, the fermionic and the scalar gauge couplings. One concrete occurrence of such amplitude is the stop pair production at lepton colliders in the minimal supersymmetric standard model, where there is a photon exchange in the s-channel and $g_\psi$ and $g_S$ are the electromagnetic gauge couplings to leptons and stops~\cite{MSSMFR}.

There is also the possibility of having a P-wave-dominated process in models where the structure of the interactions is such that we have contributions to $\psi \bar \psi \to SS$ that come from diagrams involving the one-loop exchange of vector bosons. 
For instance, if we consider a theory where a gauge field $V_\mu$ couples minimally to $\psi$ (with vector and axial couplings) while $S$ is a neutral scalar component that has a $SVV$ coupling induced by a symmetry breaking mechanism, then we can write a box diagram (see Fig. \ref{fig:case3}) which has the following amplitude:
\begin{align}
{\cal M}^{\rm P-wave}_0(\psi  \bar{\psi} \to S S) =  F_0(q,p)\bar{v}(q_{2})\slashed{p}\,u(q_1) + 
G_0(q,p) \bar{v}(q_{2})\gamma_{5} \slashed{p}\, u(q_1)\, ,
\end{align}
where $F_0$ and $G_0$ are one-loop structure functions. In general, these form factors are given in terms of complicated functions of momenta and masses but assume a simple form in the limit of heavy $V_\mu$. The form of this amplitude is similar to the one obtained by considering the leading box diagrams in Higgs pair production at electron colliders, which involve $W$ and $Z$ bosons in the loop \cite{diHiggs_Vermaseren}.
In general, the $F_0$ and $G_0$ form factors have additional $p$ dependence which makes the enhancement calculation more complicated compared with the case of having form factors that are independent of $p$, as we highlight in the next section.

In this work we calculate the Sommerfeld enhancement for both S-wave and P-wave processes, focusing on two concrete examples where the leading order amplitude is generated by an effective $\psi\bar\psi SS$ interaction. More specifically, for the S-wave case, we use as a leading order amplitude the one in Eq.~(\ref{eq:ampS}) computed in the limit of heavy $A$, while for the P-wave case we use Eq.~(\ref{eq:ampP}) as the leading order amplitude in the limit of heavy $V$. Examining  Eq.~(\ref{eq:ampS}) and Eq.~(\ref{eq:ampP}) we see that the calculation of the enhancement is actually independent of the presence of the mediator propagator in the s-channel and we can safely work in the limit where we integrate out this intermediate particle. The only difference is that using the renormalizable models one would have to consider $A/V\to SS$ as the initial process for the computation of the enhancement, while in the effective field theory limit we are working directly with the $\psi\bar\psi SS$ interaction.
In general, working with non-renormalizable $\psi\bar\psi SS$ interactions can introduce additional problems. For instance, the $\psi\bar\psi \to SS$ cross-section could be divergent already when considering one single exchange of the $\phi$ particles in the final state, in a way that this quantity is no longer a prediction of the theory~\cite{nonreno1,nonreno2} and needs to be renormalized by itself. This is not the case for the interactions we are considering in this paper since there are no additional factors of momenta in the effective vertex.
\footnote{However, there could be a problem when considering a generic effective operator which might introduce higher powers of $p$ momenta in the $\psi\bar\psi SS$ vertex. In UV complete models this is not a problem because the renormalization occurs as usual.}

\section{The enhancement factor ${\cal S}(E)$} \label{sec:Scalc}

\begin{figure}
\unitlength = 0.6mm
\centering
    \includegraphics{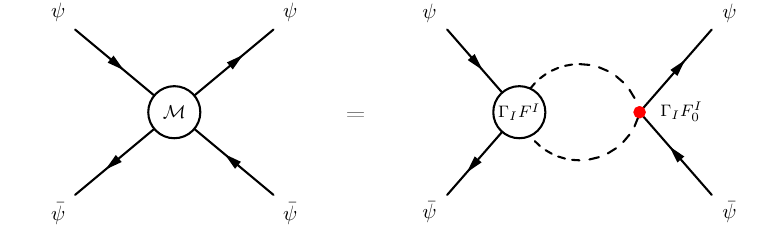}
\caption{ Recursion relation for  $\bar{\psi} \psi \to \bar{\psi} \psi$ at all orders in the exchange of the light particles (contained in the effective $\Gamma$ vertex).  }
\label{fig:optical}
\end{figure}

The infinite series of ladder diagrams, representing the exchange of an increasing number of $\phi$ particles in the final state, can be re-summed by solving a recursion relation for the non-perturbative form factor $F^{I}(q,p)$. This recursion relation is shown diagrammatically in Fig.~\ref{fig:recursion}, and the corresponding equation is given by:
\begin{figure}
    \scalebox{0.92}{\includegraphics{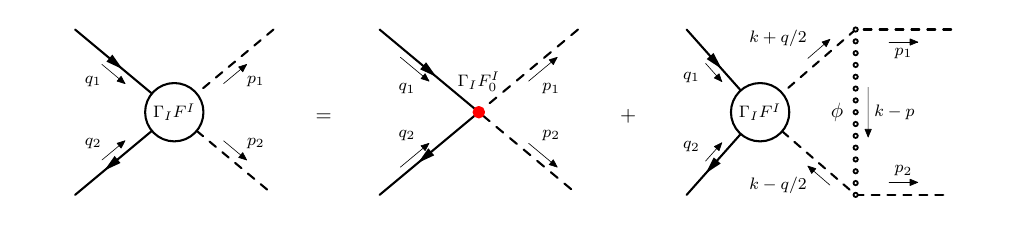}}
\caption{ Recursion relation for the leading operator in the $\psi \bar{\psi} \to S S$ process. The blob represents the leading interaction which in general can be momentum dependent. }
\label{fig:recursion}
\end{figure}
\begin{align}
\label{eq:rec_rel}
F^{I}(q,p)&=& F_0^{I}(q,p)+(i\kappa)^{2}\int \frac{\dd[4]{k}}{(2\pi)^{4}}\frac{i}{(k+\frac{q}{2})^{2}-m_{S}^{2}+im_{S}\Gamma_{S}}\frac{i}{(k-\frac{q}{2})^{2}-m_{S}^{2}+im_{S}\Gamma_{S}}\nn\\
&& \frac{i}{(k-p)^{2}-m_{\phi}^{2}}F^{I}(q,k) \, .
\end{align}
In the non-relativistic limit, we have that 
\begin{align}
 q\to(2 m_S +E, \vec{0})\qquad {\rm and} \qquad p\to (0,\vec{p}) \, ,  
\end{align}
where $E$ is the non-relativistic energy of the final state system. Using these momenta approximations and performing the $k_0$ integral, the recursion relation in Eq. \eqref{eq:rec_rel} reduces to (see Appendix \ref{ap:nonrel} for the derivation):
\begin{align} \label{eq:recursiongeneral}
F^{I}(E,\vec{p}) = F_0^{I}(E,\vec{p}) - \frac{\kappa^{2}}{4m_{S}^{2}}\int \frac{\dd[3]{k}}{(2\pi)^{3}} \frac{1}{E+i\Gamma_{S} - \frac{\vec{k}^{2}}{m_{S}}} \frac{1}{\left(\vec{k}-\vec{p} \right)^{2} + m_{\phi}^{2}}F^{I}(E,\vec{k}) \, .
\end{align}
Let us define the following function:
\begin{align}\label{gfmomspace}
\tilde G^{I}(z,\vec{p}) =  -\frac{1}{z - \frac{\vec{p}^{2}}{m_{S}}} F^{I}(E,\vec{p}) \, ,
\end{align}
where $z=E+i \Gamma_{S}$. With the above definition, we can write the recursion relation in Eq.~\eqref{eq:recursiongeneral} as:
\begin{align} 
\left(\frac{\vec{p}^{2}}{m_{S}} - z \right) \tilde G^{I}(z,\vec{p}) = F_0^{I}(E,\vec{p}) + \int \frac{\dd[3]{k}}{(2\pi)^{3}} \tilde V(\vec{k}-\vec{p}) \tilde G^{I}(z,\vec{k}) \, .
\end{align}
The quantity
\be \label{eq:yukpotft}
\tilde V(\vec{p})=-\frac{4\pi \alpha_{\phi}}{\vec{p}^{2} + m_{\phi}^{2}}
\ee
is the Fourier transform of the Yukawa potential induced by the exchange of $\phi$ 
\be \label{eq:yukpot}
V(\vec{r})=-\alpha_{\phi}\frac{e^{-m_\phi r}}{r},
\ee 
where $r=|\vec{r}|$ and $\alpha_{\phi} = \frac{\kappa^{2}}{16\pi m_{S}^{2}}$.
At this point it is useful to define the position space representation of $\tilde G^{I}(z,\vec{p})$ as follows:
\begin{align}
  G^{I}(z,\vec{r})=\int \frac{\dd[3]{p}}{(2\pi)^{3}}\, \tilde G^{I}(z,\vec{p})\,e^{i\vec{p}\cdot\vec{r}} \,.
\end{align}
Now we need to find the connection between $G^{I}(z,\vec{r})$ defined above and the $\psi\bar{\psi}  \to S S$ cross-section. In order to do so, we consider the amplitude of the 2-to-2 scattering process $\psi\bar{\psi}  \to \psi \bar{\psi}$, which is shown in Figure~\ref{fig:optical}, and compute the total cross-section for $\psi\bar{\psi}  \to S S$ by applying the optical theorem.
We can write the amplitude ${\cal M}(\psi\bar{\psi}  \to \psi \bar{\psi})$ as follows:
\begin{align}
 {\cal M}(\psi\bar{\psi}  \to \psi \bar{\psi}) = \bar{v}(q_{2})\Gamma_{I}u(q_{1})\bar{u}(q_{1})\Gamma_{J}v(q_{2}) I^{IJ}(q)   \, ,
\end{align}
where $I^{IJ}(q)$ is given by:
\begin{align}
  I^{IJ}(q) = - i \int \frac{\dd[4]{k}}{(2\pi)^{4}} F^{I}(q,k) \frac{1}{(k+\frac{q}{2})^{2}-m_{S}^{2}+im_{S}\Gamma_{S}} \frac{1}{(k-\frac{q}{2})^{2}-m_{S}^{2}+im_{S}\Gamma_{S}} F_{0}^{J}(q,k) \, .  
\end{align}
In the non-relativistic limit, the quantity $I^{IJ}(q)$ can be written in terms of the leading order form factor $F_0^J$ and the function $\tilde G^{I}(z,\vec{k})$ defined in Eq.~\eqref{gfmomspace}. We have:
\begin{align} \label{eq:generalOPT}
  I^{IJ}(E) = \frac{1}{4m_{S}^{2}} \int \frac{\dd[3]{k}}{(2\pi)^{3}}F_{0}^{J}(E,\vec{k}) \tilde G^{I}(z,\vec{k})
  =\frac{1}{4m_{S}^{2}}\mathcal{O}^{J} G^{I}(z,\vec{r})\Bigg\vert_{\vec{r}=0}  \, .  
\end{align}
The quantity $I^{IJ}(E)$ is just the Fourier transform of $F_{0}^{J}(E,\vec{k}) \tilde G^{I}(z,\vec{k})$ evaluated at $\vec{r}=0$. Notice that, after performing the Fourier transform,  the form factor $F_0^J(E,\vec{k})$ becomes a differential operator in position space, that we denoted by $\mathcal{O}^{J}$, which acts on $G^{I}(z,\vec{r})$. 
This quantity is in general divergent\footnote{The Fourier transform at $\vec{r}\neq 0$ acts as a regulator for the divergent quantity $I^{IJ}(E)$. In this case, the divergences are represented as singular terms for $\vec{r}\to 0$.}: the divergences can be real and imaginary. However, only imaginary divergences are problematic for the computation of the enhancement, as we will see in the following.

Let us compute the total cross section for $\psi\bar{\psi}  \to S S$ by applying the optical theorem:
\bea\label{opticaltsigma}
s \sigma (\bar{\psi} \psi \rightarrow S S)  &=&   \Im \sum_{\text{spins}} \mathcal{M}(\bar{\psi} \psi \rightarrow \bar{\psi} \psi)
 =  \Im \sum_{\text{spins}} \bar{v}(q_{2})\Gamma_{I}u(q_{1})\bar{u}(q_{1})\Gamma_{J}v(q_{2}) I^{IJ}(q) \nn\\
\eea
In the non-relativistic limit, we can write
\be \label{opticaltsigma2}
\sigma^{\rm NR} (\bar{\psi} \psi \rightarrow S S)  = \frac{1}{s}  g_{IJ}\Im I^{IJ}(E)
\ee
where $g_{IJ}=\Tr(\Gamma_{I}\slashed{q_1}\Gamma_{J}\slashed{q_2})$ is a real quantity.  We should point out that there are subtleties in using the optical theorem with unstable particles in internal lines. We discuss this issue in Appendix \ref{ap:OPT}, where we show that the usual Cutkosky rules apply with good approximation when the widths are narrow compared to the particle's mass.  
Using Eq.~\eqref{eq:generalOPT} and Eq.~\eqref{opticaltsigma2} we can write the Sommerfeld enhancement factor as follows:
\begin{align} \label{eq:general}
{\cal S}(E) = \frac{\sigma^{\rm NR}(\psi  \bar{\psi} \to S  S)}{\sigma^{\rm NR}_{0}(\psi  \bar{\psi} \to S  S)} = \frac{g_{IJ}\Im \mathcal{O}^{J} G^{I}(z,\vec{r})\Big\vert_{\vec{r}=0}}{g_{IJ}\Im \mathcal{O}^{J} G_{0}^{I}(z,\vec{r})\Big\vert_{\vec{r}=0}}    \, , 
\end{align}
where $G_{0}^{I}(z,\vec{r})$ is the Fourier transform of the function defined in Eq.~\eqref{gfmomspace} computed in the free case, namely for $\alpha_{\phi} = 0$. The specific form of the enhancement factor defined in Eq.~\eqref{eq:general} is computed in the following sections for different types of interactions (S-wave and P-wave) and depending on the form of the interaction, the quantity $I^{IJ}(E)\sim \mathcal{O}^{J} G^{I}(z,\vec{r})\Big\vert_{\vec{r}=0} $ presents imaginary divergences which need to be renormalized to provide a finite result.

\section{S-wave Sommerfeld enhancement}\label{sec:warm}
In this section, we study the enhancement of an S-wave process. The divergences appearing in the computation of $I^{IJ}(E)$ in Eq.~\eqref{eq:generalOPT} are real~\cite{Peskin, toppair, topFadin} even when the width of $S$ is non-zero (unstable final-state particles). Therefore, the computation of the total cross-section in Eq.~\eqref{opticaltsigma2} is unaffected by the presence of divergent terms. Here we compute those divergences in any case because they help us understand the origin of imaginary divergences occurring in the P-wave case that directly affect the enhancement calculation and is discussed in the next section.

\begin{figure}
\centering
\begin{minipage}[t]{0.4\textwidth}
    \begin{subfigure}{\textwidth}
        \centering
        \includegraphics{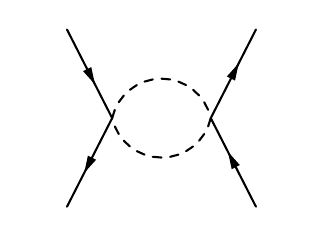}
        \caption{}
        \label{fig:I0}
    \end{subfigure}
\end{minipage}
\,
\begin{minipage}[t]{0.4\textwidth}
    \begin{subfigure}{\textwidth}
        \centering
        \includegraphics{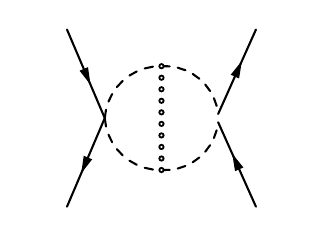}
        \caption{}
        \label{fig:I1}
    \end{subfigure}
\end{minipage}
\caption{ Feynman diagrams corresponding to the first two terms in the expansion of $I(E)$ in powers of $\alpha_\phi$.}
\label{fig:loops}
\end{figure}

Let us consider an effective dimension-5 operator ($\bar{\psi} \psi SS$) that gives the following S-wave leading order amplitude for $\psi\bar\psi\to SS$:

\begin{align}\label{eq:swavel}
\mathcal{M}_{0}^{\text{S-wave}} = \mathcal{\lambda}_{0}\bar{v}(q_{2}) u(q_{1}) \, ,
\end{align}
where $\lambda_0$ is a real dimensionful coupling constant. The same amplidude~\footnote{As already said, the Sommerfeld enhancement computed with Eq.~\eqref{eq:swavel} is identical to the one computed with Eq.~\eqref{eq:ampS}.} can be derived from Eq.~\eqref{eq:ampS} in the limit of very heavy $A$ and in this case $\lambda_0= -y_{\psi A}\kappa_{SA}/m_{A}^{2}$. By comparing Eq.~\eqref{eq:swavel} with the most general parametrization of the amplitude in Eq.~\eqref{eq:ampbase} we have that 
\be 
\Gamma_I=\mathbf{1}\qquad\qquad{\rm and}\qquad\qquad F_0^I(q,p)=\lambda_0\,.
\ee
Using these definitions together with Eq.~\eqref{eq:recursiongeneral}, we can write the following recursion relation for the non-perturbative form factor $F(E,\vec{p})$ in the non-relativistic limit:
\be\label{eq:13}
F(E,\vec{p})= \mathcal{\lambda}_{0} + \int \frac{\dd[3]{k}}{(2\pi)^{3}}\tilde{V}(\vec{k}-\vec{p})F(E,\vec{k})\frac{1}{z - \frac{\vec{k}^{2}}{m_{S}}} \, , 
\ee
where $\tilde{V}(\vec{k}-\vec{p})$ has been defined in Eq.~\eqref{eq:yukpotft}. At this point, it is convenient to define the Green's function in momentum space as follows\footnote{In this case the $F_{0}^{I}=\lambda_0$ is constant and we can divide the equation by it to simplify the calculation.}:
\begin{align}\label{eq:green}
\tilde{G}(z,\vec{k})= -\frac{1}{z - \frac{\vec{k}^{2}}{m_{S}}}\frac{F(E,\vec{k})}{\mathcal{\lambda}_{0}} \, ,
\end{align}
such that the Fourier transform of Eq.~\eqref{eq:13} becomes the familiar Schr\"odinger equation:
\begin{align} \label{eq:usual_shcro}
\left[- \frac{\nabla^{2}}{m_{S}}- z +V(\vec{r})\right]G(z,\vec{r}) = \delta^{3}(\vec{r}) \, .
\end{align}
where $V(\vec{r})$  is the Yukawa potential defined in Eq.~\eqref{eq:yukpot}. The function $G(z,\vec{r})$ is the Fourier transform of Eq.~\eqref{eq:green} and can be interpreted as the standard Schr\"odinger Green's function $G(z,\vec{r},\vec{r}')$, evaluated at $\vec{r}'=0$.
The next step is to use the optical theorem to find the relation between the cross-section and the imaginary part of the Green's function. We can start from Eq.~\eqref{eq:generalOPT} and write the integral as:
\begin{align} \label{eq:generalOPTSwave}
I(E) = \frac{\lambda_{0}^{2}}{4m_{S}^{2}}\int \frac{\dd[3]{k}}{(2\pi)^{3}} \tilde{G}(z,\vec{k})= \frac{\lambda_{0}^{2}}{4m_{S}^{2}} G(z,\vec{r}) \Big\vert_{\vec{r}=0} \, .
\end{align}
Notice that the integral $I(E)$ is simply given by the Green's function $G(z,\vec{r})$ evaluated at $\vec{r}=0$.
Therefore, the optical theorem in Eq.~\eqref{opticaltsigma2} can be written as: 
\be
\sigma(\bar{\psi} \psi \rightarrow S S) =\frac{1}{s}\Tr(\slashed{q}_{2}\slashed{q}_{1})\Im I(E)=
\frac{1}{s} \Tr(\slashed{q}_{2}\slashed{q}_{1})\frac{\lambda_{0}^{2}}{4m_{S}^{2}}\Im G(z,\vec{r}) \Big\vert_{\vec{r}=0}\, 
\ee
The Sommerfeld enhancement factor in Eq.~\eqref{eq:general} is given by:
\begin{align}\label{eq:enhancementSwave}
{\cal S}(E) = \frac{\Im G(z,\vec{r})\vert_{\vec{r}=0}}{\Im G_{0}(z,\vec{r})\vert_{\vec{r}=0}}.    
\end{align}
From now on we will consider the limit $m_{\phi} \rightarrow 0$. In this limit, the Yukawa potential in Eq.~\eqref{eq:yukpot} becomes the Coulomb potential and Eq.~\eqref{eq:usual_shcro} admits an analytic solution~\cite{coulombGF}. To identify the divergent terms of $I(E)\sim G(z,\vec{r})\Big |_{\vec{r}=0}$ we take the analytic Coulomb Green's function $G(z,\vec{r},\vec{r}')$~\cite{coulombGF} evaluated at $\vec{r}'=0$ and expand it around $\vec{r}=0$. Therefore we get:
\begin{align}\label{eq:swavediv}
I^{\text{div}}(E) =
\frac{\lambda_{0}^{2}}{4m_{S}^{2}} \left(\frac{m_{S}}{4\pi r} + \frac{m_{S}^{2}\alpha_{\phi}}{4\pi} \log r  \right)  \Bigg\vert_{\vec{r}=0} \, .
\end{align}
We can see that the divergent terms of the Coulomb Green's function at $\vec{r}=0$ are real and they do not affect the computation of the enhancement factor in Eq.~\eqref{eq:enhancementSwave} which involves only the imaginary part of $G(z,\vec{r})\vert_{\vec{r}=0}$. Nevertheless, it is very instructive to analyze the origin of these divergences. In order to do this, let us use Eq.~\eqref{eq:13} and Eq.~\eqref{eq:green} to solve Eq.~\eqref{eq:generalOPTSwave} by expanding the recursion relation for $F(E,\vec{k})$ in powers of $\alpha_\phi$. We can write
\be \label{eq:iesexp}
I(E) = I_{0}(E)+ I_{1}(E) + \ldots + I_n(E) +\ldots \, ,
\ee
where $I_n(E)=I^{\rm finite}_n(E)+I^{\rm div}_n(E)$ contains terms of order $\alpha_\phi^n$. 
Here we compute the first two terms of the series in Eq.~\eqref{eq:iesexp} and identify the divergent pieces. These quantities, $I_{0}(E)$ and $I_{1}(E)$, represent the non-relativistic limit of the loop integrals shown in Fig.~\ref{fig:I0} and Fig.~\ref{fig:I1}, respectively. 
The first integral is
\be 
I_{0}(E) = \frac{\lambda_{0}^{2}}{4m_{S}^{2}}  \int \frac{\dd[3]{k}}{(2\pi)^{3}}  \frac{1}{\frac{\vec{k}^{2}}{m_{S}}-z} 
=\frac{\lambda_{0}^{2}}{4m_{S}^{2}} G_0(z,\vec{r})\Bigg\vert_{\vec{r}=0} \, ,
\ee
where $G_0(z,\vec{r})$ is the free Green's function which is given by
\be \label{eq:freegreenf}
G_{0}(z,\vec{r}) = \frac{m_{S}}{4\pi r} e^{i\sqrt{m_{S}z} r} \, .
\ee 
Expanding around $\vec{r}=0$ we identify the following divergent contribution
\be 
I_{0}^{\text{div}}(E)=\frac{\lambda_{0}^{2}}{4m_{S}^{2}} \left(  \frac{m_{S}}{4\pi r} \right)  \Bigg\vert_{\vec{r}=0} \, ,
\ee
which coincides with the first term of Eq.~\eqref{eq:swavediv}. 
The second integral of the expansion is 
\be 
I_{1}(E) = -\frac{\lambda_{0}^{2}}{4m_{S}^{2}}  \int \frac{\dd[3]{k}}{(2\pi)^{3}}  \frac{1}{\frac{\vec{k}^{2}}{m_{S}}-z}
\int \frac{\dd[3]{k'}}{(2\pi)^{3}}\frac{4\pi\alpha_\phi}{(\vec{k'}-\vec{k})^2}\frac{1}{ \frac{\vec{k'}^{2}}{m_{S}}-z} \ .
\ee
The position space representation can be done by integrating one of the momenta and doing the Fourier transform of what survives in the large momenta region. Expanding around $\vec{r}=0$ we identify the following divergent contribution (see Appendix~\ref{ap:swaveONE}):
\be
I_{1}^{\text{div}}(E) = \frac{\lambda_{0}^{2}}{4m_{S}^{2}} \frac{ m_{S}^{2} \alpha_{\phi}}{4\pi}\log r ,
\ee
which coincides with the second term of Eq.~\eqref{eq:swavediv}.  All the $I_n(E)$ integrals, with $n \geq 2$, are finite. Therefore, in the ladder approximation, we have that the UV divergences appearing in the computation of ${\cal M}(\psi\bar\psi\to\psi\bar\psi)$ in the non-relativistic limit are two-loop exact, namely $I^{\text{div}}(E)=I_0^{\text{div}}(E)+I_1^{\text{div}}(E)$. The $\bar{\psi} \psi \rightarrow \bar{\psi} \psi$ amplitude can be simply renormalized by introducing the following counterterm
\be \label{eq:swavecounterterm}
 \bar{v}_{s}(q_{2})u_{r}(q_{1})\bar{u}_{r}(q'_{1}) v_{r}(q'_{2})\delta I_{CT} \, ,
\ee 
with $\delta I_{CT}$ equal to $-I^{\text{div}}(E)$ given by Eq.~\eqref{eq:swavediv}. Notice that this counterterm looks non-renormalizable because we considered a non-renormalizable interaction to start with. On the other hand, in a renormalizable model, where there is another particle in the s-channel that mediates the interaction between $\psi$ and $S$, the counterterm would enter in the renormalization of that particle's self-energy.  

 The counterterm in Eq.~\eqref{eq:swavecounterterm} cancels all the divergences appearing in the non-relativistic limit in the ladder approximation in the $\psi\bar\psi\to\psi\bar\psi$ process, which is directly related to $G(z,\vec{r}) \Big\vert_{\vec{r}=0}$. We obtained this finite contribution without having to consider the non-resonant contributions. For this process, since this is two-loop exact we would need to compute at least the divergent contribution of a three-loop diagram in order to cancel out this divergence. This cancelation would not get rid of all the UV divergences, because the theory has intrinsic UV divergences which can only be dealt with renormalization. 

With this approach, we can obtain the finite resonant contribution without any ambiguities. It is possible to understand the role of this non-relativistic renormalization in dealing with non-resonant processes since we can always imagine a scenario where we have a much heavier particle in the self-energy of $S$ that should be able to be integrated out. This result is trivial for the S-wave process since these divergences do not appear in the optical theorem, but becomes non-trivial for P-wave processes as we show in the next section.

Notice that we are using a position space regularization scheme, which is not the one usually implemented in standard quantum field theory calculations. If one wants to compute, in addition to the enhancement, other observables of the theory, then the same regularization scheme needs to be implemented. This means that one would need to compute the Green's function in dimensional regularization \cite{Eiras:1999xx,dimGREEN1,dimGREEN2} or use the position space regularization scheme for the computation of the observables.

\section{P-wave Sommerfeld enhancement} \label{sec:pwave}
In the previous section, we saw that no imaginary divergences are affecting the computation of the S-wave enhancement. On the other hand, for the P-wave case, the situation is different and imaginary divergences appear. The Sommerfeld enhancement in P-wave dominated processes was studied in \cite{axialpV,pwaveTH,stopFadin}. There are different forms of P-wave-dominated amplitudes one can write down but all of them have the property of being proportional to the velocity of the final state system in the region close to the threshold. 

In this section we consider an effective dimension-6 operator ($\bar{\psi}\gamma^{\mu}  \psi S\partial_{\mu} S$)that gives the following P-wave leading order amplitude for $\psi\bar\psi\to SS$:
\begin{align}\label{eq:pwavel}
\mathcal{M}_{0}^{\text{P-wave}} = F_{0}\bar{v}(q_{2})\slashed{p} u(q_{1}) \, ,
\end{align}
where $F_0$ is a real dimensionful coupling constant. We choose this form of interaction because it is the most simple example of a P-wave process and can occur in different models. For instance, the same amplitude can be derived from Eq.~\eqref{eq:ampP} in the limit of very heavy $V$ and in this case $F_0= -g_{\psi}g_{S}/m_{V}^{2}$. 
By comparing Eq.~\eqref{eq:pwavel} with the most general parameterization of the amplitude in Eq.~\eqref{eq:ampbase} we have that \be 
\Gamma_I=\gamma^{\mu}\qquad\qquad{\rm and}\qquad\qquad F_{0}^{I}(q,p)=p_{\mu}F_{0} \, .
\ee
Using these definitions together with Eq.~\eqref{eq:recursiongeneral}, we can write the following recursion relation for the non-perturbative form factor $F^i(z,\vec{p})$ \footnote{Because we are using effective interactions, it could be that  $F^i(z,\vec{p})$ is also divergent. This divergence would spoil the predictability of the process and is something to be careful of in a general model. In our case, the amplitude is finite and we do not have this specific problem.} in the non-relativistic limit:
\begin{align}\label{eq:55}
F^{i}(z,\vec{p})&=p^{i}F_{0}+ \int \frac{\dd[3]{k}}{(2\pi)^{3}}\tilde{V}(\vec{k}-\vec{p})F^{i}(z,\vec{k})\frac{1}{z - \frac{\vec{k}^{2}}{m_{S}}} \,, 
\end{align}
where $\tilde{V}(\vec{k}-\vec{p})$ has been defined in Eq.~\eqref{eq:yukpotft}. In analogy to the S-wave case, it is convenient to define:
\begin{align}\label{eq:vecgreen}
\tilde{G}^{i}(z,\vec{k})= -\frac{1}{z - \frac{\vec{k}^{2}}{m_{S}}}\frac{F^{i}(z,\vec{k})}{F_{0}}\equiv k^{i}\tilde{G}(z,\vec{k}) \,,
\end{align}
where in the last equality we have used the fact that $F^{i}(z,\vec{p})\sim p^i$, since $p^{i}$ is the only vector we have at our disposal to construct a covariant expression for $F^{i}(z,\vec{p})$. 
With this definition, the position space representation of Eq.~\eqref{eq:55} becomes: 
\begin{align} \label{eq:deriva0}
\left[ - \frac{\nabla^{2}}{m_S}- z +V(\vec{r}) \right]\partial_{i}G(z,\vec{r}) = \partial_{i}\delta^{3}(\vec{r}) \,,
\end{align}
where $V(\vec{r})$ is the Yukawa potential defined in Eq.~\eqref{eq:yukpot} and $G(z,\vec{r})$ is the Fourier transform of $\tilde{G}(z,\vec{k})$ defined in Eq.~\eqref{eq:vecgreen}. Let us show that $G(z,\vec{r})$ can be taken to be the standard Schr\"odinger Green's function $G(z,\vec{r},\vec{r}')$ that solves 
\begin{align} \label{eq:deriva2}
\left[ - \frac{\nabla^{2}}{m_{S}}- z +V(r)\right]G(z,\vec{r},\vec{r}') =\delta^{3}(\vec{r}-\vec{r}') \, ,
\end{align}
evaluated at $\vec{r}'=0$. In order to do this, let us take Eq.~\eqref{eq:deriva2} and act with $- \partial_{i}'$ on both sides:
\begin{align} \label{eq:deriva1}
\left[ - \frac{\nabla^{2}}{m_{S}}- z +V(r)\right]\left(-\partial_{i}'G(z,\vec{r},\vec{r}') \right) = -\partial_{i}'\delta^{3}(\vec{r}-\vec{r}') \, .
\end{align}
Using the fact that $\partial'_{i}G(z,\vec{r},\vec{r}')=-\partial_{i}G(z,\vec{r},\vec{r}')$ and substituting this relation back into Eq.~\eqref{eq:deriva1}, it is possible to show that we recover Eq.~\eqref{eq:deriva0} after taking $\vec{r}'\to 0$. Therefore, the Green's function equation we need to solve here is the same as the one we encountered in the S-wave case. The difference appears when expressing the cross-section in terms of the Green's function, using the optical theorem. To show this let us start from Eq.~\eqref{eq:generalOPT} and write the integral as
\begin{align}\label{eq:integralpvawe}
I^{ij}(E) = \frac{F_{0}^{2}}{4m_{S}^{2}}\int \frac{\dd[3]{k}}{(2\pi)^{3}} k^{j}\tilde{G}^{i}(z,\vec{k})
= \frac{F_{0}^{2}}{4m_{S}^{2}}\int \frac{\dd[3]{k}}{(2\pi)^{3}} k^{i}k^j\tilde{G}(z,\vec{k})
\, ,
\end{align}
where in the second equality we used Eq.~\eqref{eq:vecgreen}. Thanks to the symmetry properties of the integrand, we can replace $k^ik^j\to \delta^{ij} k^2/3$ inside the second integral of Eq.~\eqref{eq:integralpvawe}. In this way we obtain:
\begin{align}\label{eq:integralpvawe2}
I^{ij}(E) = \frac{F_{0}^{2}}{12m_{S}^{2}}\delta^{ij}\int \frac{\dd[3]{k}}{(2\pi)^{3}} k^2\tilde{G}(z,\vec{k})
= -\frac{F_{0}^{2}}{12m_{S}^{2}}\delta^{ij} \nabla^2G(z,\vec{r})\Big|_{\vec{r}=0}\, .
\end{align}
Notice that the integral $I^{ij}(E)=\delta^{ij}I(E)$ is simply given in terms of the Laplacian of the Green's function that solves Eq.~\eqref{eq:deriva2} evaluated at $\vec{r}=\vec{r}'=0$. The optical theorem in Eq.~\eqref{opticaltsigma2} can be written as: 
\bea
\sigma(\bar{\psi} \psi \rightarrow S S) &=&\frac{1}{s}\Tr(\slashed{q}_{2}\gamma_i\slashed{q}_{1}\gamma_j)\delta^{ij}\Im I(E)
=-\frac{F_{0}^{2}}{12m_{S}^{2}} \frac{1}{s}\Tr(\slashed{q}_{2}\gamma_i\slashed{q}_{1}\gamma^i)\Im \nabla^2 G(z,\vec{r}) \Big\vert_{\vec{r}=0}\,. \nn\\
\eea 
In the P-wave enhancement calculation, differently than the S-wave case, the total cross section is given in terms of the Laplacian of $G(z,\vec{r})$ and the Sommerfeld enhancement factor ${\cal S}(E)$ can be written as:
\begin{align}
S(E) = \frac{\Im \nabla^{2}G(z,\vec{r}) \Big\vert_{\vec{r}=0}}{\Im \nabla^{2}G_{0}(z,\vec{r}) \Big\vert_{\vec{r}=0 }}.
\end{align}
Let us compute $I^{ij}(E)=\delta^{ij}I(E)\sim \delta^{ij} \nabla^{2}G(z,\vec{r}) \big\vert_{\vec{r}=0}$ perturbatively in $\alpha_\phi$ and identify the divergent terms. In order to do this we
use Eq.~\eqref{eq:55} and Eq.~\eqref{eq:vecgreen} to solve Eq.~\eqref{eq:integralpvawe} by expanding the recursion relation for $F^i(z,\vec{k})$ in powers of $\alpha_\phi$. We can write
\be 
I^{ij}(E) = \delta^{ij}\left(I_{0}(E)+ I_{1}(E) + \ldots + I_n(E) +\ldots\right) \, ,
\ee
where $I_n(E)$ contains terms of order $\alpha_\phi^n$. Let us consider the first two integrals of the expansion, namely $I_{0}(E)$ and $I_{1}(E)$. They are UV divergent and represent the non-relativistic limit of the loop integrals shown in Fig.~\ref{fig:I0} and Fig.~\ref{fig:I1}, respectively. The divergences of the Green function are then matched order by order in the perturbative expansion.

The first integral is:
\begin{align}
I_{0}(E) &=  \frac{F_{0}^{2}}{12m_{S}^{2}}  \int \frac{\dd[3]{k}}{(2\pi)^{3}} \vec{k}^{2} \frac{1}{\frac{\vec{k}^{2}}{m_{S}}-z} 
= -\frac{ F_{0}^{2}}{12m_{S}^{2}} \nabla^{2}G_{0}(z,\vec{r})\Bigg|_{\vec{r}=0}\,,
\end{align}
where $G_0(z,\vec{r})$ is the free Green's function given by Eq.~\eqref{eq:freegreenf}. Expanding around $\vec{r}=0$ we identify the following $1/r$ divergent contribution:
\begin{align}\label{eq:i0divpvawe}
I_{0}^{\rm div}(E)= \frac{ F_{0}^{2}}{12m_{S}^{2}} \frac{m_{S}^{2}z}{4\pi r} \Bigg|_{\vec{r}=0}\ \, .
\end{align}
Notice that, in this case, the imaginary part of the $1/r$ divergence is non-zero for finite width, namely for $\Gamma_S\neq 0$. The second integral of the expansion is:
\be 
I_{1}  = -\frac{F_{0}^{2}\alpha_{\phi}\pi}{3m_{S}^{2}}  \int \frac{\dd[3]{k}}{(2\pi)^{3}}\frac{\dd[3]{l}}{(2\pi)^{3}} (k\cdot l) \frac{1}{z - \frac{\vec{k}^{2}}{m_{S}}}\frac{1}{(\vec{k}-\vec{l})^{2}+m_{\phi}^{2}}\frac{1}{z - \frac{\vec{l}^{2}}{m_{S}}}\, .
\ee
This integral is UV divergent and it can be regularized by taking the Fourier transforms for a generic $\vec{r}\neq 0$. Then, expanding around $\vec{r}=0$, it is possible to identify the following divergent contributions (see Appendix~\ref{ap:pwaveONE}):
\begin{align}\label{eq:i1divpvawe}
I_{1}^{\text{div}}(E) = \frac{F_{0}^{2}}{12m_{S}^{2}} \alpha_{\phi}\left(- \frac{m_{S}^{2}}{8\pi} \frac{1}{r^{2}} + \frac{m_{S}^{3}z}{4\pi} \log r \right)\Bigg|_{\vec{r}=0}
\end{align}
Notice that, in this case, we have a new real $1/r^2$ divergence and the imaginary part of the $\log r$ divergence is non-zero for finite width, namely for  $\Gamma_S\neq 0$. Contrary to the S-wave case, we have that the UV divergences appearing in the computation of ${\cal M}(\psi\bar\psi\to\psi\bar\psi)$ in the non-relativistic limit are not two-loop exact. Indeed, in the P-wave case, additional divergent contributions are coming from higher loops. However, it is possible to show that these divergent terms can only be of the form  $1/r^{2}$, $1/r$, and $\log r$, and no new $r$ dependence appears. To do this, let us consider the $n+1$-loop expression:
\begin{align}\nonumber
I_{n} (E) &= -\frac{F_{0}^{2} \pi^{n}\alpha_{\phi}^{n}}{3m_{S}^{2}}  \int \frac{\dd[3]{k_{1}}}{(2\pi)^{3}}\frac{\dd[3]{k_{2}}}{(2\pi)^{3}}\ldots \frac{\dd[3]{k_{n}}}{(2\pi)^{3}} k_{1}\cdot k_{n} \frac{1}{z - \frac{\vec{k}_{1}^{2}}{m_{S}}}\frac{1}{(\vec{k}_{1}-\vec{k}_{2})^{2}+m_{\phi}^{2}}\frac{1}{z - \frac{\vec{k}^{2}_{2}}{m_{S}}} \\ &\frac{1}{(\vec{k}_{2}-\vec{k}_{3})^{2}+m_{\phi}^{2}}\frac{1}{z - \frac{\vec{k}^{2}_{3}}{m_{S}}}\ldots \frac{1}{(\vec{k}_{n-1}-\vec{k}_{n})^{2}+m_{\phi}^{2}}\frac{1}{z - \frac{\vec{k}^{2}_{n}}{m_{S}}}   \, .
\end{align}
By inspecting the integrand, one can see that the divergences come from the $k_{1}$ and $k_{n}$ integration, every other integral in between is finite. Therefore, the divergences have the same functional form as the one-loop $I_{0}(E)$ and two-loop $I_{1}(E)$ case. We can write
\be \label{eq:pwavediv}
I^{\text{div}}(E) =  \frac{A}{r^2} +  \frac{B}{r} + C \log r
\ee 
where $A$, $B$ and $C$ are complex coefficients that can be written as a series in $\alpha_\phi$.  Since $I(E)\sim \nabla^2 G(z,\vec{r})\Big |_{\vec{r}=0}$, the all-order expression of these coefficients can be obtained by considering the Laplacian of the Coulomb Green's function $G(z,\vec{r},\vec{r}')$~\cite{coulombGF} evaluated at $\vec{r}'=0$ and expand it around $\vec{r}=0$. The terms in Eq.~\eqref{eq:i0divpvawe} and Eq.~\eqref{eq:i1divpvawe} are recovered by expanding the full result to ${\cal O}(\alpha_\phi)$. The $\bar{\psi} \psi \rightarrow \bar{\psi} \psi$ amplitude can be simply renormalized by introducing the following counterterm
\be \label{eq:pwavecounterterm}
 \bar{v}_{s}(q_{2})\gamma_{i}u_{r}(q_{1})\bar{u}_{r}(q'_{1})\gamma_{i} v_{r}(q'_{2})\delta I_{CT} \, ,
\ee 
with $\delta I_{CT}$ equal to $-I^{\text{div}}(E)$ given by Eq.~\eqref{eq:pwavediv}. The counterterm in Eq.~\eqref{eq:pwavecounterterm} cancels all the divergences appearing in the non-relativistic limit in the ladder approximation in the $\psi\bar\psi\to\psi\bar\psi$ process, which is directly related to $\nabla^2 G(z,\vec{r}) \Big\vert_{\vec{r}=0}$. In order to compute the Sommerfeld enhancement, one can start directly with the Laplacian of Green's function and remove all the divergent imaginary terms there. In the P-wave case, these divergences have an imaginary part (proportional to the width $\Gamma_S$) and therefore this procedure obtains a finite contribution for the optical theorem.

\begin{figure}[t!]
 \resizebox{0.40\linewidth}{!}{\includegraphics{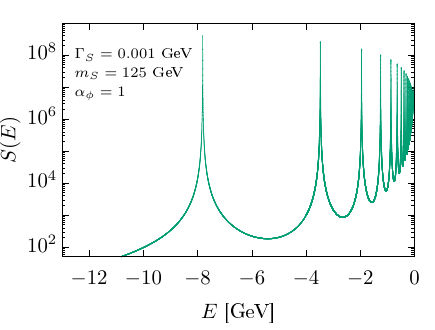}}
  \resizebox{0.40\linewidth}{!}{\includegraphics{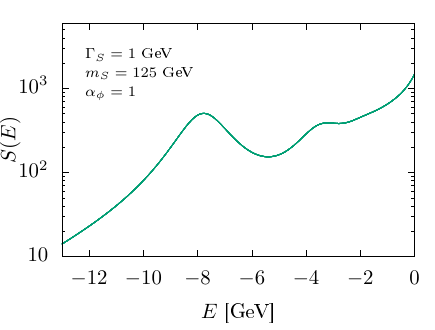}}
   \resizebox{0.40\linewidth}{!}{\includegraphics{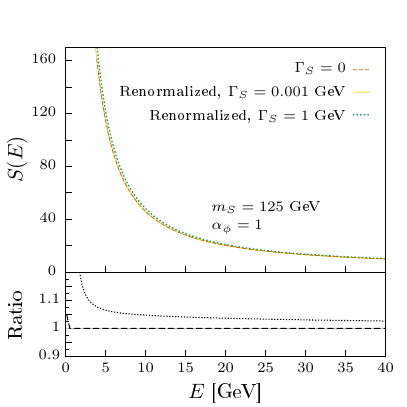}}

  \caption{P-wave dominated \label{fig:somme10} Sommerfeld enhancement factor $S(E)$ behavior in the below-threshold region, computed using $\alpha_{\phi} =1$, $m_{S} = 125 $ GeV and $\Gamma_{S} = 0.001$ GeV (top left) or $\Gamma_{S} = 1$ GeV (top right). In the bottom figure, the comparison between the behavior of the Sommerfeld enhancement in the above threshold region is presented, computed using $\alpha_{\phi} =1$, $m_{S} = 125 $ GeV and two non-zero width values, namely $\Gamma_{S} = 0.001$ GeV and $\Gamma_{S} = 1$ GeV, and the literature result~\cite{Iengo:2009ni} obtained with $\Gamma_{S} = 0$. The inset of the bottom figure shows the ratio between the renormalized enhancement and the results from \cite{Iengo:2009ni} (dashes), and the ratio between the renormalized enhancements for $\Gamma_{S} = 1$ GeV and  $\Gamma_{S} = 0$ (dots). }
\end{figure}
As an example of the application of this method, we perform the computation of the Sommerfeld enhancement for some specific values of the parameters as a function of the energy $E$. In the leftmost Figure~\ref{fig:somme10} we show the behavior of the Sommerfeld enhancement factor as a function of the energy $E<0$ (below threshold region),  computed for $\alpha_{\phi}=1$, $m_{S} = 125 $ GeV and $\Gamma_{S} = 0.001$ GeV. In the rightmost Figure~\ref{fig:somme10} we show the same behavior computed for $\alpha_{\phi}=1$, $m_{S} = 125 $ GeV and $\Gamma_{S} = 1$ GeV. From these two plots, we can see that the finite $S$ width has the effect of smearing out the delta function spikes associated with the energy levels of the $SS$ bound state and this effect is more visible when increasing the width value. We verified that in the case of larger widths, the calculation breaks down as expected due to large unitarity violations, as discussed in Appendix~\ref{ap:OPT}.  Notice that the region which is most sensitive to the width is below the threshold, which is also the region where non-resonant contributions can be significant. This means that, apart from the location of the poles, it is not possible to say anything else meaningful without computing the finite non-resonant piece.

\begin{figure}[t!]
 \resizebox{0.44\linewidth}{!}{\includegraphics{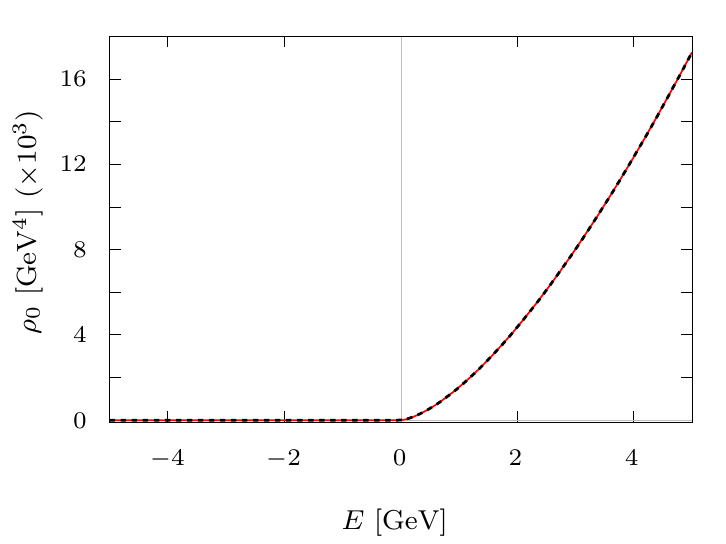}}
 \resizebox{0.49\linewidth}{!}{\includegraphics{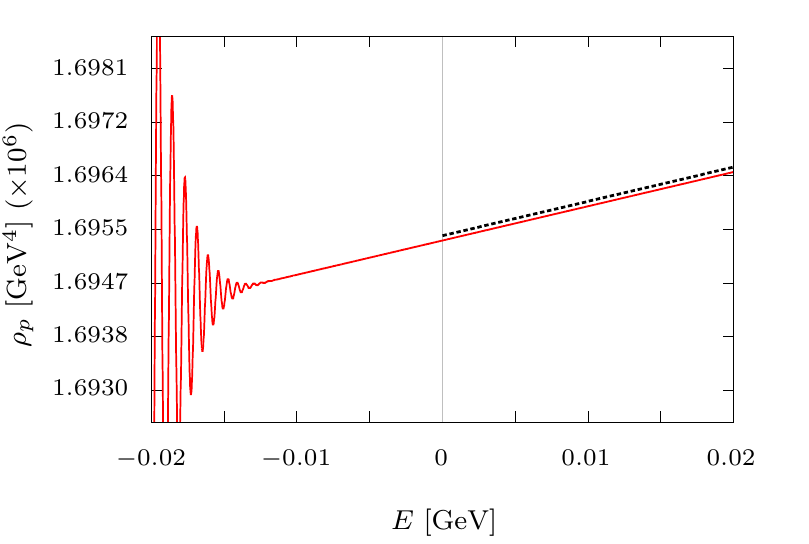}}

  \caption{P-wave dominated \label{fig:spectral} spectral function, computed using $\alpha_{\phi} = 1$, $m_{S} = 125 $ GeV and $F_{0} = 1$.  On the left, we have the spectral function for the free case with the same labeling. On the right, we have the Coulomb spectral function where the continuous red line has $\Gamma_{S} = 0.001$ GeV and the dashed black line has zero width. We focus on the region close to the threshold showing the continuous behavior from the inclusion of the width. }
\end{figure}
In the bottom Figure~\ref{fig:somme10} we show the behavior of the Sommerfeld enhancement factor as a function of the energy $E>0$ (above threshold region),  computed for $\alpha_{\phi}=1$, $m_{S} = 125 $ GeV and two different non-zero values of $\Gamma_{S}$ which are taken to be $0.001$~GeV and $1$~GeV, and we compare it with the $\Gamma_{S} = 0$ literature result~\cite{Iengo:2009ni}. For $E>0$, we have that the finite width effects are small compared to $\Gamma_{S} = 0$ case (less than 10\%).

We can also explore the spectral function directly, which we define as $\rho_{p} = \Im \nabla^{2}G(z,\vec{r}) \Big\vert_{\vec{r}=0}$. The spectral function is the relevant quantity in the computation of thermally averaged 
cross sections for dark matter production that can also be Sommerfeld-enhanced in P-wave processes \cite{KIM2019469}. In Figure~\ref{fig:spectral} we have both the free and Coulomb interaction P-wave spectral function for $\Gamma_{S}= 0.001$~GeV and zero width. The spectral function is continuous crossing $E=0$ for finite width. As we decrease $\Gamma_{S}$ the peaks become more pronounced and get closer to $E=0$. The plot cannot display the behavior for negative energy and zero width, as the distribution transforms into a summation of delta functions positioned at each peak\cite{stopFadin}.

\section{Discussion} \label{sec:discuss}

Now that we have presented our method of estimation for the  Sommerfeld enhancement,  it is worth discussing what is the difference between the approach that we propose here and what was used in the literature before. In previous works on the topic, the divergence was noticed for the case where there are unstable particles in P-wave processes. There were different attempts at solving this problem. In \cite{stopFadin}, it was discussed how the divergence is unavoidable in the non-relativistic limit, and an arbitrary cutoff around the scale of the stop mass was adopted. In \cite{pwaveTH}, it was discussed how the uncertainties in choosing the right cutoff for the stop pair production are of the same order of magnitude as higher-order QCD corrections. In this case, there is no clear scale for what the cutoff should be and the reduced mass was used as this sets the scale of the theory. In this approach, the enhancement is acknowledged to be UV sensitive and the physical cutoff shows the dependence on new physics. Near the cutoff, it is expected for order one correction to appear from the new physics contribution. This is different from the approach that we introduce in this work. 

Afterward, it was realized that this imaginary divergence is an artifact of separating the resonant and non-resonant contributions of the self-energy~\cite{Beneke:2003xh,Beneke:2004km,Jantzen:2013gpa}. However, depending on the application, the finite non-resonant contributions can be truly sub-dominant and safely ignored~\cite{Slatyer:2009vg, Beneke:2022rjv}. In principle, we cannot ignore the divergent non-resonant contribution which is used to make the calculation finite. Our goal is to obtain a finite all-orders resonant contribution that includes the width effects without having to perform non-resonant perturbative calculations. Our prescription provides a simple solution to this problem. Provided that the width is small such that the optical theorem can reliably be applied in the case of unstable particles~\cite{Denner:2014zga,Donoghue:2019fcb}, we can compute the finite resonant enhancement near threshold without the need to compute the non-resonant contributions. At this point, we reinforce that this prescription is an approximation, and if one is interested in precise below-threshold effects there is no way around computing the non-resonant contributions. The region in which we can neglect the non-resonant contributions is above threshold and with non-relativistic energies.

\begin{figure}[h!]
 \resizebox{0.6\linewidth}{!}{\includegraphics{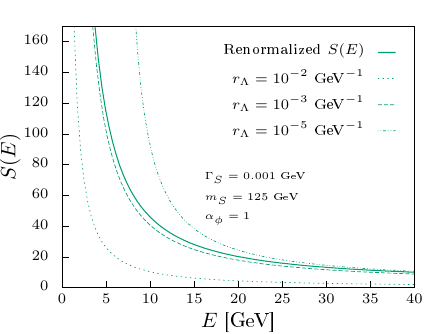}}
    \caption{ Enhancement value ($S(E)$) for the interaction $\slashed{q_{3}}$ in the above-threshold regime with $\alpha_{\phi} =1$ and $m_{S} = 125 $ GeV for  $\Gamma_{S} = 0.001$ GeV. The enhancement is presented for positive energies computed using the renormalization approach (Solid curve) and by using different values for a cutoff given by $r_{\Lambda}$ for $\Gamma_{S} = 0.001$ GeV }
\label{fig:somme2}
\end{figure}

If we were working in a model where the UV is strongly interacting, then the cutoff can have a better motivated physical meaning as being the dynamical scale itself. In the models that we study here, there is no dynamic scale. This makes it harder to guess what cutoff could give a physical result. In the position space regularization, a cutoff is defined as a minimal distance $r_{\Lambda}$. The only natural cutoff in our case is the scale at which the non-relativistic theory ceases to be a good approximation. In this case, we should expect when the distances probed are smaller than the reduced mass the approximation starts to break down. Because the divergences are of order $1/r^{2}$, there is a strong sensitivity to the scale. If there were only logarithmic divergences the sensitivity near zero would be softer. To compare the approach we develop with a cutoff procedure, we choose three different cutoffs and compare them with the method proposed here: $r_{\Lambda} = 10^{-2} \;\text{GeV}^{-1}, r_{\Lambda} = 10^{-3} \;\text{GeV}^{-1}$ and $r_{\Lambda} = 10^{-5}\; \text{GeV}^{-1}$. Since we are using the mass $m_{S} = 125 ~\text{GeV}$ we expect that the best cutoff lies around $1/125~\text{GeV}^{-1} $. We explore these cutoffs in Figure~\ref{fig:somme2} for the case where the width is $\Gamma_{S} = 0.001 $. We can see that the cutoff which is closer to the true result is the one that is closer to the $1/m_{S}$ scale. 

In the case where the width is larger, we could not find a sensible cutoff, and the behavior of the enhancement factor breaks down. This regime is unphysical because the cutoff procedure incorporates either too much of the divergent contribution or too much of the finite contribution, and thus the unitarity violation is significant and the approximation ceases to work.

We can see then that, given a reasonable choice of cutoff, it is possible to reproduce the renormalized result to a good approximation. However, the sensitivity to the cutoff makes it difficult to choose one, to begin with. In this case, it seems better to approximate the problem to zero width, in which we do not have any divergences than to guess what is the best value for the cutoff. In the approach we propose, we use the consistency of the renormalized quantum field theory to guarantee our result for the Sommerfeld enhancement is sensible and independent of UV physics. This removes the arbitrariness of choosing the cutoff, even if it is physical, and allows us to probe the small, but measurable, finite width effects.

\section{Conclusion} \label{sec:CONC}
In this work, we have proposed a method to deal with divergences that arise when computing the Sommerfeld enhancement factor in P-wave processes with unstable particles in the final state. This problem has been recognized in the literature before and previous studies showed that these divergences can be avoided by using a simple cutoff procedure.  Later it was shown that the imaginary divergences appearing in the calculations are a consequence of separating the resonant and non-resonant contributions. In this work, we show that these divergences can  also be removed, from the resonant contribution, by performing a renormalization procedure at low energies and therefore it is not necessary to use any (physical or unphysical) cutoff in the calculation. In this way, we can reliably calculate the finite all-orders resonant contribution of the Sommerfeld enhancement. We showed that for the specific P-wave process we have studied, the effects of the finite width are small (of the order of $10 \%$)for pair production above threshold.

 In the case where $\Gamma_{S}/m_{S}$ is small, there are three ways to estimate the total cross-section with different levels of precision: \\
$\bullet$ assume stable particles and perform the standard procedure, obtaining a finite result, 
which ignores both the width and the non-resonant contribution;\\
$\bullet$ include the width and perform the non-relativistic renormalization with complex coefficients, which includes the width effects and ignores the non-resonant contribution (this work); \\
$\bullet$ perform the full calculation of the resonant and non-resonant contribution, obtaining a more precise result depending on 
which order in perturbation theory one is computing. 

In this paper, we have used a specific model to implement our non-relativistic renormalization procedure. However, the procedure that we introduce here can be generalized to any kind of interaction. In approaching a different model, the first step is to verify that the 2-to-2 cross-section, which we want to enhance, is finite (and thus predictable) when considering the loop exchange of the mediator. This check can be done perturbatively and it is always true for renormalizable theories. Then, the next step is to apply the optical theorem, take the non-relativistic limit and identify which operator acts on the Green's function. From that, it is guaranteed that all the divergences are dealt with in the renormalization, and thus one should worry only about picking the finite contribution for the enhancement. A renormalization scheme must be chosen, and if other observables are computed in the same theory, they need to be calculated in the same scheme.  In the case where the Green's function is only obtained numerically, the subtraction can introduce some additional error since it will be necessary to work out what is the finite contribution to the process. This, unfortunately, is a problem even if the amount to be subtracted is known exactly order by order since numerical noise makes the cancellation not exact. 

 In recent years, the scenario where dark matter has an unstable sector is becoming more popular \cite{decayDM1,decayDM2}. This work can be useful for the calculation of enhancement of 2-to-2 cross-section in beyond the standard models with unstable sectors without calculating the perturbative non-resonant contributions, provided that the width is small.

\section*{Acknowledgments}
\noindent We would like to thank Gabriel Menezes for the helpful discussions. The work of AT and CHL is supported in part by the Natural Sciences and Engineering Research Council of Canada (NSERC).  The work of AV is supported by F.R.S.-FNRS through the IISN convention "Theory of Fundamental Interactions" (N : 4.4517.08)  and the MISU convention F.6001.19, and a CNPq Ph.D. fellowship. The work of RR is supported in part by the São Paulo Research Foundation (FAPESP) through grant \# 2021/10290-2, by a CNPq Productivity grant and by an Erna and Jakob Michael Visiting Professorship in the Department of Particle Physics and Astrophysics at the Weizmann Institute of Science, that he thanks for the hospitality.

\appendix
\section{Non-relativistic approximation of the recursion relations.}\label{ap:nonrel}
In this appendix, let us work out the non-relativistic approximations of
the recursion relations used in the paper. Starting from:
\begin{align}
F^{I}(q,q_1,p_1)&=& F_0^{I}(q,q_1,p_1)+(i\kappa)^{2}\int \frac{\dd[4]{k}}{(2\pi)^{4}}\frac{i}{(k+\frac{q}{2})^{2}-m_{h}^{2}+im_{h}\Gamma_{h}}\frac{i}{(k-\frac{q}{2})^{2}-m_{h}^{2}+im_{h}\Gamma_{h}}\nn\\
&& \frac{i}{(k-p)^{2}-m_{\phi}^{2}}F^{I}(q,q_1,k) \, .
\end{align}
We then consider only contributions linear in the time-like direction inside the integral:
\begin{align}
\frac{i}{(k+\frac{q}{2})^{2}-m_{S}^{2}+im_{S}\Gamma_{S}} &\rightarrow \frac{i}{2m_{S} \left( E/2+k_{0} - \frac{\vec{k}^{2}}{2m_{S}} + i \Gamma_{S}/2\right) } \, , \\
\frac{i}{(k-\frac{q}{2})^{2}-m_{S}^{2}+im_{S}\Gamma_{S}} &\rightarrow \frac{i}{2m_{S} \left( E/2-k_{0} - \frac{\vec{k}^{2}}{2m_{S}} + i \Gamma_{S}/2\right) }  \, , \\
\frac{i}{(k-p)^{2}-m_{\phi}^{2}} &\rightarrow \frac{-i}{ (\vec{k}-\vec{p})^{2}+m_{\phi}^{2} }  \, .
\end{align}
The recursion relation in this limit is then:
\begin{align} \nonumber
F^{I}(E,\vec{p})&=& F_0^{I}(E,\vec{p})-\frac{(i\kappa)^{2}}{4m_{S}^{2}}\int \frac{\dd[4]{k}}{(2\pi)^{4}} \frac{i}{E/2+i\Gamma_{S}/2+k_{0}-\frac{\vec{k}^{2}}{2m_{S}}}\frac{i}{E/2+i\Gamma_{S}/2-k_{0}-\frac{\vec{k}^{2}}{2m_{S}}} \\
&& \frac{i}{\left(\vec{k}-\vec{p} \right)^{2} + m_{\phi}^{2}}F^{I}(E,\vec{k}) \, ,
\end{align}
where we used $q=(2 m_S +E, \vec{0})$, with $E$ being the energy above the threshold for the production of the S particle pair (negative $E$ corresponds to scattering below threshold). In the non-relativistic limit, we have $p^0=0$ and thus we can perform the $k^{0}$ integral:
\begin{align}
\int \frac{\dd{k^{0}}}{2\pi}\frac{i}{2m_{S} \left( E/2-k_{0} - \frac{\vec{k}^{2}}{2m_{S}} + i \Gamma_{S}/2\right) } \frac{i}{2m_{S}\left( E/2 +k^{0}-\frac{\vec{k}^{2}}{2m_{S}} +i\Gamma_{S}/2\right)} = \frac{i}{4m^{2}_{S}} \frac{1}{\left( E +i \Gamma_{S} - \frac{\vec{k}^{2}}{m_{S}}\right)}\, .
\end{align}
With this we recover Eq.\eqref{eq:recursiongeneral}:
\begin{align}
F^{I}(E,\vec{p}) = F_0^{I}(E,\vec{p}) - \frac{\kappa^{2}}{4m_{S}^{2}}\int \frac{\dd[3]{k}}{(2\pi)^{3}} \frac{1}{E+i\Gamma_{S} - \frac{\vec{k}^{2}}{m_{S}}} \frac{1}{\left(\vec{k}-\vec{p} \right)^{2} + m_{\phi}^{2}}F^{I}(E,\vec{k}) \, .
\end{align}

\section{Optical theorem for unstable particles} \label{ap:OPT}
\begin{figure}[h!]
 \resizebox{0.9\linewidth}{!}{\includegraphics{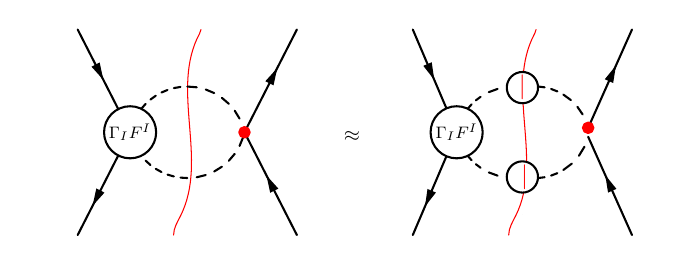}}
\caption{\label{fig:narrowdith}The cut through the unstable internal line in the narrow width approximation on the left side of the equation. On the right is the actual cut in the stable states in the case where we have this two-body decay. }
\end{figure}

In the optical theorem approach, it is also possible to treat unstable particles in any internal line. The procedure uses the complex mass scheme\cite{Denner:2014zga,complexmass1} where we change the internal propagator to have the decay width:
\begin{align}
\frac{1}{p^{2} - m^{2}} \rightarrow \frac{1}{p^{2} - m^{2} + i m \Gamma}
\end{align}
At this point, one should be careful when considering these unstable particles, especially with the unitarity of the theory. The application of Cutkosky rules does not apply directly to unstable particles \cite{VELTMAN1963186,Denner:2014zga}. However, a general relation (largest time equation) can be obtained, which is valid for both unstable and stable particles \cite{VELTMAN1963186}. Using the largest time equation, it is possible to show that any cut with unstable particles does not contribute to the scattering matrix. The physical interpretation is that no final state has those particles. Because of this, we can, in the end, use the Cutkosky rules for both unstable and stable particles, but we do not cut the unstable states\footnote{Applying Cutkosky or holomorphic cuts to unstable-particle propagators may lead to well-defined results, once they are treated properly as distributions~\cite{CUTUNSTABLE}.} \cite{Donoghue:2019fcb}. The downside of this approach is that processes that go through the unstable state end up being in higher loop order, as represented in Figure \ref{fig:narrowdith}.

The way to simplify such treatment can be achieved when the particle has a narrow width. In this case, we can treat the unstable particles just as stable particles because we separate the production from the decay. The cut generates the imaginary part of the propagator, which in the narrow width approximation has the same delta function form as the stable particle:
\begin{align}
\Im\left(\frac{1}{p^{2} - m^{2} + i m \Gamma} \right) = -\frac{m\Gamma}{(p^{2}-m^{2})^{2}+m^{2}\Gamma^{2}} \approx -\pi \delta(p^{2}-m^{2})
\end{align}

In this case, we can get some effects for the finite width while preserving the optical theorem's simplicity and unitarity. For a more in-depth discussion about this, we refer the reader to \cite{Denner:2014zga,Donoghue:2019fcb}. With this in mind, we can then use the optical theorem for states with narrow width, which is the main interest of this paper.

\section{One-loop process (two-loop optical diagram) for the S-wave contact interaction.} \label{ap:swaveONE}
The analysis for the one-loop process(two-loop optical diagram) is similar to the leading contribution:
\begin{align}
i\mathcal{M} &=  i \bar{v}_{s}(p_{2})u_{r}(p_{1})\bar{u}_{r}(p_{1}) v_{r}(p_{2}) I_{1} \\ \nonumber
I_{1} &=  -\lambda_{0}^{2}\kappa^{2} \int \frac{\dd[4]{k}}{(2\pi)^{4}}\frac{\dd[4]{l}}{(2\pi)^{4}} \frac{1}{(k+q/2)^{2}-m_{S}^{2}+im_{S}\Gamma_{S}} \frac{1}{(k-q/2)^{2}-m_{S}^{2}+im_{S}\Gamma_{S}} \\
& \frac{1}{(k-l)^{2}-m_{\phi}^{2}}\frac{1}{(l-q/2)^{2}-m_{S}^{2}+im_{S}\Gamma_{S}}\frac{1}{(l+q/2)^{2}-m_{S}^{2}+im_{S}\Gamma_{S}} \, .
\end{align}
Then, we apply the near-threshold approximation by dropping every quadratic dependence on the energy or the time component of the vectors from the denominator. In this case we drop $k_{0}^{2}$, $l_{0}^{2}$, $k_{0}l_{0}$, $k_{0}E$, $l_{0}E$ and $E^{2}$. Using this approximation, the integral in the non-relativistic limit becomes:
\begin{align} \nonumber
I_{1}  &=   -\lambda_{0}^{2}\kappa^{2}\int \frac{\dd[4]{k}}{(2\pi)^{4}}\frac{\dd[4]{l}}{(2\pi)^{4}} \frac{1}{2m_{S} (k_{0}+\frac{E+i\Gamma_{S}}{2} - \frac{\vec{k}^{2}}{2m_{S}})}\frac{1}{2m_{S} (k_{0}-\frac{E+i\Gamma_{S}}{2} - \frac{\vec{k}^{2}}{2m_{S}})} \\
&\frac{-1}{(\vec{k}-\vec{l})^{2}+m_{\phi}^{2}}\frac{1}{2m_{S} (l_{0}+\frac{E+i\Gamma_{S}}{2} - \frac{\vec{l}^{2}}{2m_{S}})}\frac{1}{2m_{S} (l_{0}-\frac{E+i\Gamma_{S}}{2} - \frac{\vec{l}^{2}}{2m_{S}})} \, .
\end{align}
The integration of $k_{0}$ and $l_{0}$ is done to give:
\begin{align} \label{eq:36}
I_{1} &= -\frac{\lambda_{0}^{2}\kappa^{2}}{16m_{S}^{4}}  \int \frac{\dd[3]{k}}{(2\pi)^{3}}\frac{\dd[3]{l}}{(2\pi)^{3}}  \frac{1}{E +i \Gamma_{S} - \frac{\vec{k}^{2}}{m_{S}}}\frac{1}{(\vec{k}-\vec{l})^{2}+m_{\phi}^{2}}\frac{1}{E +i \Gamma_{S} - \frac{\vec{l}^{2}}{m_{S}}}\, .
\end{align}

To write the divergent contribution in position space, we must first perform the following $l$ integral
\be
I_{1}^{l} = \int \frac{\dd[3]{l}}{(2\pi)^{3}} \frac{1}{(\vec{k}-\vec{l})^{2}+m_{\phi}^{2}}\frac{1}{ \frac{\vec{l}^{2}}{m_{S}}-z}\,.
\ee
Using the Feynman parametrization, we can rewrite it as:
\be 
I_{1}^{l} = m_{S}\int_{0}^{1}\dd{x} \int \frac{\dd[3]{l}}{(2\pi)^{3}} \frac{1}{(\vec{l}^{2}+\Delta^{2})^{2}}
= \frac{m_{S}}{8\pi}  \int_{0}^{1}\dd{x} \frac{1}{\sqrt{\Delta^{2}}}\,,
\ee 
where $\Delta^{2} = (1-x)(\vec{k}^{2}+m_{\phi}^{2}) -\vec{k}^{2} (1-x)^{2} -m_{S}z x$. We can perform the integral in $x$, which is well behaved in the limit when $m_{\phi} \rightarrow 0$ and write:
\begin{align}
I_{1} = -\lambda_{0}^{2} \kappa^{2}    \int \frac{\dd[3]{k}}{(2\pi)^{3}} \frac{\sqrt{-k^{2}+m_{S}z}}{64\pi m_{S}^{2}k(k^{2}-m_{S}z)^{3/2}} \sinh^{-1}\left( \frac{k}{\sqrt{-k^{2}+m_{S}z}} \right) \, .
\end{align}
We can expand for large momenta to pick up the most divergent contributions and then perform the Fourier transform:
\begin{align}
I_{1}  = -\frac{\lambda_{0}^{2}\kappa^{2}_{\phi}}{16m_{S}^{4}} \frac{m_{S}}{8\pi}  \int \frac{\dd[3]{k}}{(2\pi)^{3}} \left(m_{S}\pi\frac{1}{|\vec{k}|^{3}}   + \ldots \right)
\end{align}
Now if we perform the Fourier transform using:
\begin{align}
\mathcal{F}\left(\frac{1}{k^{3}}\right) &= -\frac{1}{2\pi^{2}} \log r \, ,
\end{align}
We can write the one-loop divergent contribution as:
\begin{align} \nonumber
I_{1}^{\text{div}} &=
-\frac{\lambda_{0}^{2}}{4m_{S}^{2}}\frac{\kappa^{2}_{\phi}}{4m_{S}^{2}} \left( -\frac{m_{S}^{2}}{16\pi^{2}} \right)  \log r  \\
 &= \frac{\lambda_{0}^{2}}{4m_{S}^{2}}\alpha_{\phi} \left( \frac{ m_{S}^{2}}{4\pi} \right)  \log r  \, .
\end{align}

\section{One-loop process (two-loop optical diagram) for the P-wave contact interaction.}\label{ap:pwaveONE}
The next-to-leading for the P-wave process can be computed as follows:
\begin{align}
i\mathcal{M} &=  i  \bar{v}_{s}(p_{2})\gamma_{\mu}u_{r}(p_{1})\bar{u}_{r}(p_{1})\gamma_{\nu} v_{r}(p_{2}) I^{\mu\nu}_{1} \\ \nonumber
I^{\mu \nu}_{1} &=  -F_{0}^{2}\kappa^{2} \int \frac{\dd[4]{k}}{(2\pi)^{4}}\frac{\dd[4]{l}}{(2\pi)^{4}} k^{\mu} l^{\nu} \frac{1}{(k+q/2)^{2}-m_{S}^{2}+im_{S}\Gamma_{S}} \frac{1}{(k-q/2)^{2}-m_{S}^{2}+im_{S}\Gamma_{S}} \nonumber \\
& \frac{1}{(k-l)^{2}-m_{\phi}^{2}}\frac{1}{(l-q/2)^{2}-m_{S}^{2}+im_{S}\Gamma_{S}}\frac{1}{(l+q/2)^{2}-m_{S}^{2}+im_{S}\Gamma_{S}} \, .
\end{align}
We can apply the near-threshold approximation by dropping every quadratic dependence on the energy or the time component of the vectors from the denominator and perform the $k_{0}$ and $l_{0}$ integrals:
\begin{align}
I^{i j}_{1}  &= -\frac{F_{0}^{2}\kappa^{2}}{16m_{S}^{4}}  \int \frac{\dd[3]{k}}{(2\pi)^{3}}\frac{\dd[3]{l}}{(2\pi)^{3}} k^{i} l^{j} \frac{1}{E +i \Gamma_{S} - \frac{\vec{k}^{2}}{m_{S}}}\frac{1}{(\vec{k}-\vec{l})^{2}+m_{\phi}^{2}}\frac{1}{E +i \Gamma_{S} - \frac{\vec{l}^{2}}{m_{S}}}\, .
\end{align}
From this limit we can see that the only tensor structure that can appear is proportional to the Euclidean metric:
\begin{align}
&I^{i j}_{1}  = \delta^{ij} \left(I_{1} \right)_{\text{NR}} \, ,\\
&I_{1}  =- \frac{F_{0}^{2}\kappa^{2}}{48m_{S}^{4}}  \int \frac{\dd[3]{k}}{(2\pi)^{3}}\frac{\dd[3]{l}}{(2\pi)^{3}} k.l \frac{1}{\frac{\vec{k}^{2}}{m_{S}}-z } \frac{1}{(\vec{k}-\vec{l})^{2}+m_{\phi}^{2}}\frac{1}{\frac{\vec{l}^{2}}{m_{S}}-z } \, .
\end{align}

To write the divergent contribution in position space we can draw a connection with the S-wave calculation. We can use the Feynman parametrization to write $I_{1}^{l}$ as:
\be 
I_{1}^{l} = \int \frac{\dd[3]{l}}{(2\pi)^{3}} \frac{k.l}{(\vec{k}-\vec{l})^{2}+m_{\phi}^{2}}\frac{1}{ \frac{\vec{l}^{2}}{m_{S}}-z} =m_{S}\int_{0}^{1}\dd{x} \int \frac{\dd[3]{l}}{(2\pi)^{3}} \frac{(1-x)k^{2}}{(\vec{l}^{2}+\Delta^{2})^{2}} = \frac{m_{S}}{8\pi}  \int_{0}^{1}\dd{x} \frac{(1-x)k^{2}}{\sqrt{\Delta^{2}}} \, , 
\ee
where $\Delta$ is the same as for the S-wave. We can perform the $x$ integral to write $I_{1}$ as: 
\begin{align}
I_{1} = F_{0}^{2} \kappa^{2}    \int \frac{\dd[3]{k}}{(2\pi)^{3}} \frac{k\sqrt{-k^{2}+m_{S}z}}{192\pi m_{S}^{2}(k^{2}-m_{S}z)^{3/2}} \sinh^{-1}\left( \frac{k}{\sqrt{-k^{2}+m_{S}z}} \right) \, .
\end{align}
We can then expand in large momenta to perform the Fourier transform of the most divergent contributions:
\begin{align}
I_{1} = \int \frac{\dd[3]{k}}{(2\pi)^{3}}  \left( -F_{0}^{2}\pi \alpha_{\phi} \frac{1}{48 k} - F_{0}^{2} \pi z \alpha_{\phi} \frac{1}{24 k^{3}} + \ldots \right) \, .
\end{align}
Performing the Fourier transform we have:
\begin{align} \nonumber
I_{1}^{\text{div}}&= -\frac{F_{0}^{2}}{12m_{S}^{2}} \frac{\kappa^{2}_{\phi}}{4m_{S}^{2}} \left( \frac{m_{S}^{2}}{32\pi^{2}} \frac{1}{r^{2}} - \frac{m_{S}^{3}z}{16\pi^{2}} \log r \right)  \\
 &=\frac{F_{0}^{2}}{12m_{S}^{2}} \alpha_{\phi}\left(- \frac{m_{S}^{2}}{8\pi} \frac{1}{r^{2}} + \frac{m_{S}^{3}z}{4\pi} \log r \right) \, .
\end{align}

\bibliography{bibREN.bib}

\end{document}